\documentclass[aps, pre, preprint]{revtex4-1}

\usepackage{amsmath}
\usepackage{amssymb}
\usepackage{mathtools}
\usepackage{bm}
\usepackage{xspace}
\usepackage{graphicx}
\usepackage{epstopdf}
\usepackage{xcolor}
\usepackage[T1]{fontenc}
\usepackage[pagewise]{lineno}
\usepackage{hyperref}
\hypersetup{
  bookmarksopen = true, 
  bookmarksnumbered = true, 
  colorlinks   = true, 
  urlcolor     = blue, 
  linkcolor    = red, 
  citecolor    = blue 
}
\newcommand*{\VC}[1]{\bm{\mathrm{#1}}}
\newcommand*{\ER}[1]{Eq.~\eqref{#1}\xspace}
\newcommand*{\FiR}[1]{Fig.~\ref{#1}\xspace}

\newcommand*{\Pero}{\mathrm{Pe}}
\newcommand*{\epdot}{\dot{\varepsilon}}
\newcommand*{\gamdot}{\dot{\gamma}}
\newcommand*{\hatS}{\hat{S}}

\newcommand{\tr}[0]{\textrm{Tr}}

\newcommand{\SD}[1]{#1}

\begin{document}

\title{Mechanistic constitutive model for wormlike micelle solutions with flow-induced 
structure formation}

\author{ Sarit Dutta}
\email{sdutta4@wisc.edu}
\author{ Michael D. Graham}
\email{mdgraham@wisc.edu}

\affiliation{ Department of Chemical \& Biological Engineering, \\
	 University of Wisconsin-Madison, 1415 Engineering Drive, Madison, WI 53706, USA }

\date{\today}

\begin{abstract} 
We present a tensor constitutive model for predicting stress and flow-induced
structure formation in dilute wormlike micellar solutions.  The micellar
solution is treated as a dilute suspension of rigid Brownian rods whose length
varies dynamically. Consistent with the mechanism presented by Turner and
Cates [J. Phys.:~Condens. Matter 4, 3719 (1992)], flow-induced alignment of
the rods is assumed to promote increase of rod length that corresponds to the
formation of flow-induced structures observed in experiments. At very high
deformation rate, hydrodynamic stresses causes the rod length to decrease.
These mechanisms are implemented in a phenomenological equation governing the
evolution of rod length, with the number density of rods appropriately
modified to ensure conservation of surfactant mass. The model leads first to
an increase in both shear and extensional viscosity as deformation rate
increases and then to a decrease at higher rates.  If the rate constant for
flow-induced rod growth is sufficiently large, the model predicts a
multivalued relation between stress and deformation rate in both shear and
uniaxial extension.  Predictions for shear and extensional flow at steady
state are in reasonable agreement with experimental results. By design, the model is
simple enough to serve as a tractable constitutive relation for computational
fluid dynamics studies.
\end{abstract}

\maketitle


\section{Introduction}

Surfactant solutions are used in a wide range applications from fracking fluids
in the oil and gas industry to heat-transfer fluids in the chemical process
industries to various personal care and cleaning products \cite{Larson1999}.
Specifically, it has long been known that dilute surfactant solutions can behave
very differently in turbulent flow than do simple fluids, leading to dramatic
reductions in energy consumption \cite{Shenoy1984}.  Particularly in Japan, use
of surfactants as drag reducing additives in water is becoming increasingly
widespread in closed-loop district heating and cooling systems for large
buildings and other facilities, resulting in substantial energy savings
\cite{Krope2010, Saeki2011}.  Surfactants that form wormlike micelles are found
to have drag reduction characteristics very similar to those of dilute polymer
solutions \cite{Ohlendorf1986, Bewersdorff1988, Gasljevic2001, Tamano2010}. 

At sufficiently high concentration, surfactants self-assemble into micelles that
can take on a variety of forms including cylindrical, or wormlike.  Wormlike
micellar surfactant solutions (WMS) display extremely complex behavior when
subjected to flow, and in particular can display dramatic increases in viscosity
that arise due to a flow-induced transition from an isotropic liquid to a
nematic-like \emph{gel} phase \cite{Wunderlich1987, Liu1996, Hu1998,
Berret2000}. In many fluids this transition is transient \cite{Berret2000} while
in others it can be irreversible \cite{Vasudevan2010}.  The basic phenomenology
underlying this transition is reasonably well-understood (see \cite{Lerouge2009}
for a comprehensive review). However, this understanding has not yet been
translated into a mathematical model (constitutive equation) that can be used to
predict the stresses and rheological state of surfactant solutions in 
complex flow systems characteristic of many engineering applications such
as~turbulent drag reduction. Here we present a simple mechanistic model for
dilute wormlike micellar solutions that can serve as a tractable constitutive
relation in computational fluid dynamics simulations. We emphasize from the
outset that, in the interest of generating a tractable constitutive model,
our approach dramatically oversimplifies many aspects of the flow-induced
structure formation problem -- we are proposing an engineering model, not an
\emph{ab initio} one. 

Over the past three decades, a large body of experimental studies involving
surfactants of different chemical compositions have unambiguously demonstrated
the existence of dramatic shear-thickening in dilute WMS  at shear rates higher
than a critical value \cite{Rehage1982, Wunderlich1987, Boltenhagen1997, Hu1998,
Berret2000, Macias2003, Dehmoune2007}.  Analogous observations have been made in
extensional flow \cite{Prudhomme1994}, but with a critical extension rate that
is smaller than the critical rate in shear.  Imaging studies by Berret and
coworkers \cite{Berret2000, Berret2002} and Pine and coworkers
\cite{Liu1996, Boltenhagen1997, Hu1998, Hu1998a} show that shear-thickening is
accompanied by formation of a highly elastic and birefringent gel-like structure
-- the  so-called \emph{flow-induced structure} (FIS).  For many cationic
systems, substantial evidence exists that at the high deformation rates
characteristic of turbulent flow, transient FIS form, leading to drag reduction
levels that are substantially higher than found in dilute polymer systems
\cite{Lin2000}. Despite its increasingly frequent application, this phenomenon
remains poorly understood at a fundamental level. 

PIV measurements in circular Couette flow \cite{Hu1998} (in which the inner
cylinder is rotating) indicate that the FIS
grows from the inner wall of the Couette cell, eventually filling the gap and
resulting in plug flow in the core with lubricating layers near the wall.
Smaller gaps shift the critical shear rate to higher values -- Pine and
coworkers \cite{Hu1998} interpret this as a consequence of the slipping of the
FIS at the walls \cite{Lerouge2009}.  However, the growth of FIS can also be
viewed in terms of development of vorticity bands, which we further describe
below \cite{Dhont2008}. The Weissenberg numbers based on the relaxation time of
the FIS phase can be extremely large ($10^{1}-10^{3}$) at the critical shear
rate, suggesting that elasticity-driven flow instabilities will be a distinct
possibility \cite{Lerouge2009}.  At even higher shear rates, shear-thinning is
ultimately observed. 

Experiments by the Pine group \cite{Hu1998, Hu1998a} in both shear rate
controlled and shear stress controlled settings have demonstrated the existence
of a multivalued or \emph{reentrant} flow curve (plot of shear stress
\emph{vs.}~shear rate).  Note that this behavior is quite distinct from the
\emph{flow curve multiplicity} found in models of entangled WMS and polymer
melts.  For \emph{dilute} solutions, multiple stresses can be found for the same
shear rate, while for \emph{entangled} solutions, multiple shear rates can
display the same shear stress. The latter multiplicity underpins the
shear-banding phenomenon \cite{Fielding2007, Olmsted1999a}, while the former,
which is the topic of the present work, can in principle lead to so-called
vorticity bands \cite{Dhont2008}.  In a shear flow experiment the shear stress
is constant in the gradient direction while the nominal shear rate is constant
along with vorticity direction. Thus constitutive behavior that exhibits
multiple stable shear stresses for the same shear rate can display bands with
different behavior along the vorticity axis, i.e.~vorticity bands. 

In dilute WMS, the multivalued stress is observed only in a
stress-controlled experiment; the corresponding behavior in a shear rate
controlled experiment manifests as a discontinuous jump in shear stress at a
particular shear rate. Reentrant behavior was also reported by Dehmoune et al
\cite{Dehmoune2007}, whereas a discontinuous jump in shear rate controlled
experiments was reported in studies by several groups \cite{Ohlendorf1986,
Bewersdorff1988, Bandyopadhyay2001}. Note that shear-thickening without
reentrant behavior has also been observed in several studies \cite{Berret1998,
Berret2000, Berret2001, Macias2003}.

Based on experimental observations, the rheological behavior of dilute WMS has
been classified into three regimes -- (\emph{i}) at low shear rates, the
solution remains essentially Newtonian, (\emph{ii}) at intermediate shear rates
shear-thickening occurs, accompanied by the formation of FIS, and (\emph{iii})
at very high shear rates shear thinning is observed, along with breakdown of FIS
\cite{Lerouge2009}. Some studies \cite{Hu1998a, Bandyopadhyay2001} indicate the
existence of four regimes, where the second regime in the former classification
is divided into two regimes -- the first corresponding to inhomogeneous
nucleation of FIS, and the second to homogeneous nucleation of FIS.

In most experiments with dilute WMC solutions, the FIS is transient and the
fluid will eventually revert to an isotropic low viscosity state with negligible
birefringence \cite{Lerouge2009}.  Recently, Shen and coworkers
\cite{Vasudevan2010} performed an experiment where dilute WMS was driven through
a microfluidic device containing a  packed bed of glass beads to create a
microfluidic packed bed (porous medium).  In such a geometry, the flow has a
very large extensional component whereas most prior experiments have been
performed in shear rheometers. Under these conditions, they found that the
solution formed an irreversible rather than a reversible gel. In this paper, we
will be concerned solely with reversible FIS.

Only a small number of studies have attempted to put forth physically motivated
yet mathematically tractable constitutive models for surfactant solutions.  In
the context of thixotropic fluids, Bautista, Manero, Puig, and coworkers
\cite{Bautista1999} introduced a model (now known as the BMP
model) whose variants have been used in predictions of shear-thickening in
dilute WMS \cite{Manero2007, Landazuri2016} as well as shear-banding in
entangled WMS \cite{Manero2007, Bautista2007, GarciaSandoval2012}. The
generalized BMP model, which can capture shear-thickening as well as reentrant
behavior \cite{Landazuri2016}, couples a standard upper convected Maxwell
equation for the micellar contribution to the stress with an evolution equation
for the \emph{fluidity} (inverse viscosity) of the solution. The viscosity and
modulus that appear in the Maxwell equation are functions of fluidity.  The
evolution equation for fluidity consists of a term that captures spontaneous
breakdown of micellar structure, and a second term that accounts for the buildup
of structure due to work done on the fluid. Moreover, the rate constants for
structure buildup and breakdown are made to depend on the stress. However, the
use of the Maxwell equation in the BMP model is problematic in extensional flow,
as the steady state extensional viscosity will diverge at a finite extension rate.

A phenomenological model that qualitatively captures shear-thickening and
reentrant behavior was proposed by Goveas and Pine \cite{Goveas1999}. Without
appealing to any microstructural mechanism, their model considers an insoluble
gel phase to form when the local stress in the solution exceeds a critical
value. Beyond this critical stress, the evolution of the gel phase is governed
by the rate of gel formation that depends on the local stress, and a constant
rate of gel destruction. 

Shear-thickening due to onset of instability above a
critical shear rate was investigated by Barentin and Liu \cite{Barentin2001}.
According to their model, this instability arises from electrostatic attractions
between individual micelles, which fosters the growth of micellar bundles that
eventually form networks, giving rise to shear-thickening.

An important class of models is based on population dynamics, where the
evolution of sub-populations of micelles of different lengths are tracked, and
the material properties are calculated based on the interactions within each
sub-population as well as between different sub-populations. A particularly
successful model in this class was developed by Vasquez, McKinley, and Cook
\cite{Vasquez2007}, albeit for explaining the rheology of entangled WMS.  While
their model considered only two sub-populations, very good qualitative agreement
with some rheological data for entangled micellar solutions have been obtained,
while other experimental observations, such as second-normal stresses and
complex transients are still not accessible \cite{Pipe2010}. The model does
display shear-banding \cite{Zhou2008,Zhou2010} as well as the rupture phenomenon
experimentally observed in uniaxial extensional flow \cite{Cromer2009}.  Models
based on three sub-populations -- long micelles, short micelles, and a gel phase
have also been proposed \cite{Kang2016}. Depending on appropriate parameter
values, this model predicts shear-thickening as well as shear-banding and
spatial inhomogeneity. However, the presence of a large number of empirical
constants in the model poses significant difficulty for experimental
validation.

A relatively simple theoretical model of a dilute solution of rodlike micelles
that qualitatively captures the key experimental observation of a flow-induced
gelation transition has been proposed by  Cates and Turner (CT)
\cite{Cates1990,Turner1992}. In their model, rods (\emph{i}) align with the flow
and diffuse rotationally, (\emph{ii}) can react end-to-end to form longer rods,
and (\emph{iii}) can spontaneously break up into shorter rods. The reaction to
form longer rods only happens when rods are collinear.  The model takes the form
of a master equation for the evolution of the probability distribution function
for rod orientation and length; in the absence of reactions, each rod would
evolve the same way that a rigid rod evolves in dilute solution and the model
reduces to the standard Fokker-Planck equation for that case.  In elongational
flow, which strongly aligns the rods, CT  find that the reaction to form long
rods dominates so that eventually the rod  length diverges leading to a
\emph{gelation transition}. In shear, a sharp transition is absent but the mean
micellar length still becomes much larger than in the absence of flow.  This is
extremely important work, as it clearly indicates the dominant physics of
structure formation in these systems.  Nevertheless, it leaves unanswered many
questions that are important if we are to be able to quantitatively reproduce
experimental observations or make predictions for flow systems in practical
applications.  In particular, CT do not write an evolution equation for the
stress tensor -- a constitutive equation. While micelles have been modeled as
rigid rods in several past theoretical approaches, notably by Cates and Turner
\cite{Cates1990, Turner1992}, we are not aware of any model that provides a
constitutive relation predicting shear-thickening with a reentrant phase
diagram. 

To perform direct numerical simulations for dilute surfactant solutions, a
constitutive equation to relate the local stress with the local
deformation gradient is required.  To our knowledge, no closed form continuum-level
constitutive model for FIS-exhibiting dilute surfactant solutions has been
developed that uniformly treats both shear and extensional flow.  It is
important to note that the inherent complexity of the numerical algorithms
required for solving the coupled nonlinear partial differential equations
resulting from discretization of the Navier-Stokes equations in a general domain
restricts the complexity of the constitutive relation. This is so because only
the simplest models are tractable with the current level of computational
capability, within reasonable bounds of time and resources. For example, in case
of turbulent flow in dilute polymer solutions, the most widely used constitutive
model is FENE-P \cite{Bird1987}, although numerous models exist that are much
more detailed.  Indeed, studies have shown that the FENE-P model works
remarkably well in capturing key experimental observations in viscoelastic
turbulence \cite{Graham2014}.  In this paper, we present a constitutive model for
dilute WMS that is of similar complexity as FENE-P for dilute polymers and
evaluate its predictions in simple shear and uniaxial extensional flows.


\section{Model description}

\begin{figure}
    \includegraphics[width=\linewidth]{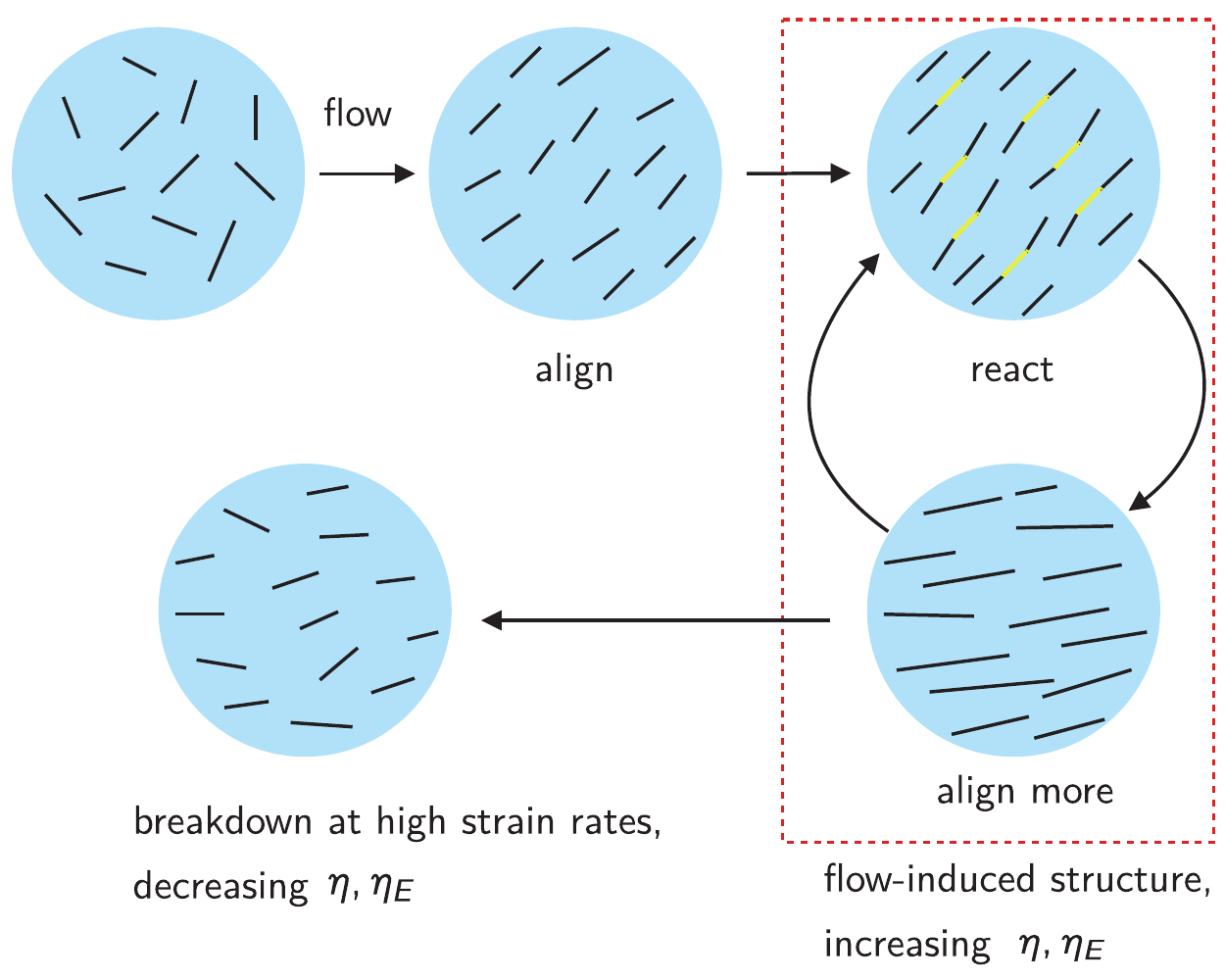}
    \caption{\label{fig:rrm} Schematic showing the mechanism of flow-induced
    structure formation according to RRM, exhibiting initial shear
    (extension)-thickening and subsequent shear (extension)-thinning behavior.}
\end{figure}

We begin by providing a qualitative description of our model, which we call the
\emph{reactive rod model} (RRM).  A schematic of the mechanisms involved is
shown in \FiR{fig:rrm}.  We consider a dilute solution of wormlike micelles as a
suspension of rigid Brownian rods. In the absence of external flow, the
orientation of the rods remains isotropic.  Since the solution is dilute, the
rods do not have steric or hydrodynamic interactions with each other. 
 
Physically, a micellar solution is polydisperse with the length following an
exponential distribution \cite{Larson1999}, but in order to make analytical
progress we will assume that a single length is sufficient to characterize the
solution at all times. Such a description has been used in prior studies of
coagulation of spherical particles to describe particle size \cite{Bremer1995}.
We emphasize from the outset that our approach dramatically oversimplifies many
aspects of the flow-induced structure formation problem in the interest of
generating a constitutive model that is tractable for computational fluid
dynamics applications. 

When an external flow field is imposed, individual rods preferentially align in
the flow direction -- the degree of alignment being governed by a balance
between rotational  diffusion and convection. We further assume that the rods
can react, such that the reaction rate increases with alignment to generate
longer rods.  This mechanism is based on experimental observations that micelles
are found to grow in length in presence of flow. As rotational diffusion of the
longer rods is slower compared to shorter rods, they align more strongly in flow
and  keep growing.  Eventually this process generates a collection of very long
rods -- a flow-induced structure that exhibits shear-thickening. However, when
the rods grow sufficiently long, they can no longer withstand hydrodynamic
stresses; they break down, leading to shear-thinning. Note that we do not
specify the microscopic mechanisms of rod coalescence or dissociation -- they are
simply accounted for by introducing phenomenological expressions that capture
key experimental observations. 

\subsection{Theory of rigid rods}

Before presenting the RRM in detail, we review some basic results for the
rheology of dilute suspension of rigid Brownian rods. Consider a dilute
suspension of rigid rods with number density $n_0$. Let each rod be of length $L_0$ and
radius $b$. The orientation of a rod is described by a unit vector $\VC{u}$.
The solvent is assumed to be Newtonian with viscosity $\eta_s$.  The suspension
is subject to a homogeneous flow $\VC{v}$ with transpose velocity gradient $\VC{K} = \nabla\VC{v}^\top$. (In Cartesian tensor notation $K_{ij}=\partial v_i/\partial x_j$.)

In homogeneous flow, the stress in a suspension depends solely on rotational
motion.  The rotational diffusion coefficient of a rod is \cite{Doi1986}
\begin{equation} \label{eq:D_r}
    D_{r,0} = \frac{3k_BT}{\pi \eta_s L_0^3}\ln\left(\frac{L_0}{2b}\right),
\end{equation}
where $k_B$ is the Boltzmann constant and $T$ is the temperature.  The average
orientation is represented by the orientation tensor $\VC{S}$, defined as the
second moment of $\VC{u}$, i.e.~$\VC{S} = \langle\VC{u}\VC{u}\rangle$, where  angle 
brackets represent ensemble average.
Starting from a Smoluchowski equation for rotational motion, the time evolution
of the orientation tensor $\VC{S}$ can be written as \cite{Doi1986}
\begin{equation} \label{eq:dSdt}
    \frac{\mathrm{d} \VC{S}}{\mathrm{d}t}  = 
    -6 D_{r,0} \left(\VC{S} - \frac{1}{3}\VC{I}\right)  
    + \VC{K}\cdot \VC{S}^{\top} + \VC{S}\cdot \VC{K}^{\top}
    - 2 \VC{K} \mathbin{:} \langle \VC{u} \VC{u} \VC{u} \VC{u}\rangle.
\end{equation}
The double dot product is defined as follows: 
$\VC{A}\mathbin{:}\VC{B}=\tr\left(\VC{A}\cdot\VC{B}^\top\right)$. 

The stress $\bm{\tau}$  in a dilute suspension of rods can be written as a sum
of the stress due to the solvent and an extra stress tensor $\bm{\tau}^p$ that
accounts for the contribution due to the presence of the rods. Thus
\begin{equation} \label{eq:tau}
    \bm{\tau} = 2 \eta_s \VC{D} + \bm{\tau}^p,
\end{equation}
where 
\begin{equation} \label{eq:taup}
    \bm{\tau}^p = 3n_0k_BT \left(\VC{S} - \frac{1}{3}\VC{I}\right)
    + \frac{n_0 k_BT}{2 D_{r,0}} \VC{K}:\langle \VC{u} \VC{u} \VC{u} \VC{u}\rangle
\end{equation}
and the rate of deformation tensor 
$\VC{D} = \left(\VC{K} +  \VC{K}^{\top}\right)/2$.

To proceed
analytically, a closure approximation for the fourth moment 
$\langle \VC{u} \VC{u} \VC{u} \VC{u}\rangle$ is necessary.  We use an expression
due to Dhont and Briels \cite{Dhont2003} that reproduces
physically reasonable rheological behavior over a very large range of shear and 
extension rates.  The approximation is as follows:
\begin{equation} \label{eq:DBC}
    \VC{K}:\langle \VC{u} \VC{u} \VC{u} \VC{u}\rangle \approx
    \frac{1}{5}\left[\VC{S}\cdot \VC{D} + \VC{D}\cdot \VC{S}
        -\VC{S} \cdot \VC{S} \cdot \VC{D}
        -\VC{D} \cdot \VC{S} \cdot \VC{S}
        + 2 \VC{S} \cdot \VC{D} \cdot \VC{S}
        + 3 \left(\VC{S}\mathbin{:}\VC{D}\right) \VC{S} 
        \right].
\end{equation}
Using \ER{eq:dSdt} -- \ER{eq:DBC}, the stress of the
suspension can be calculated for an arbitrary flow. The
results for steady state cases are obtained by setting the time derivatives to
zero.  It is known that such a suspension of rods exhibits shear-thinning and
extension-thickening \cite{Bird1987}.

\subsection{Reactive rod model (RRM)}

Now we consider the case where the rods can change length in response to flow.
In addition to the notations introduced earlier, let $n$ and $L$ represent the
number density and rod length at any time $t$ after the initiation of flow. In
contrast to $n$ and $L$, we take the radius of the rods $b$ to remain constant
at all times.  In accordance with the picture presented by Turner and Cates
\cite{Turner1992}, we assume that growth of rods increases with alignment.  From
conservation of surfactant mass (total micelle length) we have $n = n_0L_0/L$ at
all times.  Since $D_{r,0}$ is the diffusion coefficient based on rod length
$L_0$ and $D_r$ is that based on rod length $L$, from \ER{eq:D_r} we have
\begin{equation} \label{eq:Dr}
    \frac{D_r}{D_{r,0}} = \frac{1}{{L^*}^3}\left(\frac{\ln L^* + m}{m}\right),
\end{equation}
where $L^* = L/L_0$ is the rod length normalized with the initial length and $m
= \ln \left[L_0/\left(2b\right)\right]$ is a constant that serves as a measure
of  the initial aspect ratio of the rods.  

Accounting for the variation of $n$, $L$ and $D_r$, \ER{eq:dSdt} for the time
evolution of the orientation tensor becomes
\begin{equation} \label{eq:dSdt-rrm}
    \frac{\mathrm{d} \VC{S}}{\mathrm{d}t}  = 
    -6 D_r \left(\VC{S} -\frac{1}{3} \VC{I}\right)
    + \VC{K}\cdot \VC{S}^{\top} + \VC{S}\cdot \VC{K}^{\top}
    - 2\VC{K} \mathbin{:} \langle \VC{u} \VC{u} \VC{u} \VC{u}\rangle.
\end{equation}
and \ER{eq:taup} for the stress becomes
\begin{equation} \label{eq:taup-rrm}
    \bm{\tau}^p = 3nk_BT \left(\VC{S} - \frac{1}{3}\VC{I}\right)
    + \frac{n k_BT}{2D_r} \VC{K}:\langle \VC{u} \VC{u} \VC{u} \VC{u}\rangle.
\end{equation}

Further, we define a nondimensional time $t^*=D_{r,0}t$ and a P{\'e}clet number
$\Pero = \gamdot/D_{r,0}$ in shear flow and $\Pero = \epdot/D_{r,0}$ in
extensional flow. In addition, we introduce a scalar orientational order
parameter  
\begin{equation} \label{eq:hatS}
    \hatS = \sqrt{\frac{3}{2}\left(\hat{\VC{S}}\mathbin{:}\hat{\VC{S}}\right)},
\end{equation}
where $\hat{\VC{S}} = \VC{S} - \frac{1}{3}\VC{I}$ is the traceless part of
$\VC{S}$.
When the rods are completely aligned, $\hatS = 1$ whereas for isotropic
conditions (e.g.~at equilibrium) $\hatS = 0$.

Now we address the evolution of the rod length. In general we will write 
\begin{equation}\label{eq:dLdt0}
	    \frac{\mathrm{d}L^*}{\mathrm{d}t^*} = R_a+R_s,
\end{equation}
where $R_a$ represents the rate of alignment-induced growth and $R_s$ represents the rate
of spontaneous growth and breakdown of micelles.
We will take the alignment-induced growth term to increase linearly with the degree of alignment
of the rods, i.e.
\begin{equation}\label{eq:Ra}
	R_a=k\hat{S},
\end{equation}
where $k$ is a constant. 
The spontaneous growth and breakage rate $R_s$ is taken to be 
proportional to the deviation of the current length from its equilibrium value,
with a constant of proportionality $\lambda$. However, if the rods become long
enough, hydrodynamic stresses will break them apart. Thus at a given deformation
rate, there must be a maximum rod length that can be sustained without rupture.
We capture this idea by introducing the maximum rod length $L^*_{max}$. As $L^*$
approaches $L^*_{max}$ the breakage rate increases without bound, thus
forcing $L^*$ to decrease at high deformation rates. Since the hydrodynamic
stresses increase with deformation rate, the maximum rod length must decrease 
with $\Pero$.  We choose a very simple functional form of this decay as:
\begin{equation}\label{eq:Lmax}
    L^*_{max} = \alpha + \frac{\beta}{\Pero},
\end{equation}
where $\alpha$ and $\beta$ are model parameters.  Thus we take 
\begin{equation}\label{eq:Rs}
	R_s=\frac{\lambda}{1 - \left(\frac{L^*}{\alpha + \frac{\beta}{\Pero}}\right)^2}
            \left(1-L^*\right)
\end{equation}

Substituting \ER{eq:Ra} and \ER{eq:Rs} into \ER{eq:dLdt0}
 we have
\begin{equation} \label{eq:dLdt}
    \frac{\mathrm{d}L^*}{\mathrm{d}t^*} = 
        \frac{\lambda}{1 - \left(\frac{L^*}{\alpha + \frac{\beta}{\Pero}}\right)^2}
            \left(1-L^*\right) + k \hatS.
\end{equation}
\ER{eq:dLdt} combined with \ER{eq:hatS}, \ER{eq:D_r}, and \ER{eq:dSdt-rrm} results
in a system of ODEs that governs the time evolution of $\VC{S}$ and $L^*$.
The corresponding stress can be obtained by further substitution into
\ER{eq:taup-rrm}. We provide the specific equations for shear and uniaxial extension
below.


\subsection{Shear flow}

For simple shear flow, $\VC{v} = \left(\gamdot y, 0, 0 \right)^{\top}$.  Recall
that the rotational P\'{e}clet number in shear flow is defined as $\Pero =
\gamdot/D_{r,0}$. The scalar orientation parameter in this case is
\begin{equation}
    \hatS = \left[\frac{3}{2} \left\{
          \left(S_{xx} - \frac{1}{3}\right)^2
        + \left(S_{yy} - \frac{1}{3}\right)^2
        + \left(S_{zz} - \frac{1}{3}\right)^2
        + 2S_{xy}^2 \right\} \right]^{\frac{1}{2}},
\end{equation}
where $S_{xx} + S_{yy} + S_{zz} = 1$.  Substituting the closure relation
\ER{eq:DBC} into \ER{eq:dSdt-rrm} and after rearrangement we have
\begin{equation}
\begin{aligned}
    \frac{\partial S_{xx}}{\partial t^*} &= 
    -\frac{6}{{L^*}^3} \left(\frac{\ln L^* + m}{m}\right)\left(S_{xx}-\frac{1}{3}\right)
    + 2 \Pero S_{xy} - \frac{2}{5} \Pero S_{xy} \left(1 + 4 S_{xx} - S_{yy}\right)\\
    \frac{\partial S_{yy}}{\partial t^*} &= 
    -\frac{6}{{L^*}^3} \left(\frac{\ln L^* + m}{m}\right)\left(S_{yy}-\frac{1}{3}\right)
    - \frac{2}{5} \Pero S_{xy} \left(1 + 4 S_{yy} - S_{xx}\right)\\
    \frac{\partial S_{zz}}{\partial t^*} &= 
    -\frac{6}{{L^*}^3} \left(\frac{\ln L^* + m}{m}\right)\left(S_{zz}-\frac{1}{3}\right)
    - \frac{6}{5} \Pero S_{xy}S_{zz}\\
    \frac{\partial S_{xy}}{\partial t^*} &=
    -\frac{6}{{L^*}^3} \left(\frac{\ln L^* + m}{m}\right) S_{xy} 
    + \Pero S_{yy}
    - \frac{\Pero}{5}\left[6 S_{xy}^2 + S_{xx} + S_{yy} - \left(S_{xx}-S_{yy}\right)^2\right].
\end{aligned}
\end{equation}
Similar steps lead to the expressions for the stress tensor:
\begin{equation}
\begin{aligned}
    \frac{\tau^p_{xx}}{3n_0k_BT} &= \frac{1}{L^*}\left(S_{xx}-\frac{1}{3}\right)
    + \frac{m\Pero {L^*}^2}{30\left(\ln L^* + m\right) } S_{xy} \left(1 + 4 S_{xx} -S_{yy}\right) \\
    \frac{\tau^p_{yy}}{3n_0k_BT} &= \frac{1}{L^*}\left(S_{yy}-\frac{1}{3}\right)
    + \frac{m\Pero {L^*}^2}{30\left(\ln L^* + m\right)} S_{xy} \left(1 + 4 S_{yy} -S_{xx}\right) \\
    \frac{\tau^p_{zz}}{3n_0k_BT} &= \frac{1}{L^*}\left(S_{zz}-\frac{1}{3}\right)
    + \frac{m\Pero {L^*}^2}{10\left(\ln L^* + m\right)} S_{xy} S_{zz} \\
    \frac{\tau^p_{xy}}{3n_0k_BT} &= \frac{S_{xy}}{L^*} 
    + \frac{m\Pero {L^*}^2}{60\left(\ln L^* + m\right)}  
        \left[ 6S_{xy}^2 + S_{xx} + S_{yy} - \left(S_{xx}-S_{yy}\right)^2\right]
\end{aligned}
\end{equation}

\subsection{Uniaxial extensional flow}

For uniaxial extensional flow, the velocity field 
$\VC{v} = \left(-\frac{\epdot}{2}, -\frac{\epdot}{2}, \epdot \right)^{\top}$
and the scalar orientation parameter
\begin{equation}
    \hatS = \left[\frac{3}{2} \left\{
          \left(S_{xx} - \frac{1}{3}\right)^2
        + \left(S_{yy} - \frac{1}{3}\right)^2
        + \left(S_{zz} - \frac{1}{3}\right)^2
         \right\} \right]^{\frac{1}{2}},
\end{equation}
where symmetry and the unit trace condition dictates that $S_{xx} = S_{yy} = \frac{1-S_{zz}}{2}$.
Recall that in extensional flow, the rotational P\'{e}clet number is defined
as $\Pero = \epdot/D_{r,0}$. Substituting the expression for the velocity field
into the closure relation \ER{eq:DBC}, and from \ER{eq:dSdt-rrm} we have
\begin{align}
    \frac{\partial S_{xx}}{\partial t^*} &= 
    -\frac{6}{{L^*}^3} \left(\frac{\ln L^* + m}{m}\right)\left(S_{xx}-\frac{1}{3}\right)
    - \Pero S_{xx} - \frac{\Pero S_{xx}}{5} \left(9 S_{zz} -5\right)\\
    \frac{\partial S_{zz}}{\partial t^*} &=
    -\frac{6}{{L^*}^3} \left(\frac{\ln L^* + m}{m}\right) \left(S_{zz} -\frac{1}{3}\right)
    + 2\Pero S_{zz} - \frac{\Pero }{5} \left(9 S_{zz}^2 + S_{zz}\right).
\end{align}
Similar manipulations lead to the component-wise expressions for the stress
tensor:
\begin{equation}
\begin{aligned}
    \frac{\tau^p_{xx}}{3n_0k_BT} &= \frac{1}{L^*}\left(S_{xx}-\frac{1}{3}\right)
        + \frac{m\Pero {L^*}^2}{60\left(\ln L^* + m\right)} 
        S_{xx} \left(9 S_{zz} - 5\right) \\
    \frac{\tau^p_{zz}}{3n_0k_BT} &= \frac{1}{L^*}\left(S_{zz}-\frac{1}{3}\right)
        + \frac{m\Pero {L^*}^2}{60\left(\ln L^* + m\right)} 
         \left(9 S_{zz}^2 + S_{zz}\right),
\end{aligned}
\end{equation}
where $\tau^p_{xx} = \tau^p_{yy}=\tau^p_{rr}$ from symmetry, where $r=\sqrt{x^2+y^2}$ is the radial coordinate.


\section{Results and discussion}

We present results for both shear and extensional flow at steady state over a
wide range of $\Pero$ (dimensionless shear or extension rate).  Solutions are
found using the multivariate root finding algorithm \emph{hybr} as implemented
in the SciPy package \cite{Jones2001}.  
For all the results shown in the following, we take $\lambda = 1$ and $m = 7$,
the latter corresponding an initial
aspect ratio of 1000. At steady state, $\lambda$ can be combined with $k$ to
form the ratio $k/\lambda$, which is equivalent to setting $\lambda = 1$ and
varying $k$.  Varying $m$ does not have a strong effect on the qualitative
nature of the results presented. Before proceeding we note that the linear
viscoelastic behavior of this model is identical to that for rigid rods of
constant length $L_0$ (see e.g.~\cite{Doi1986}): coupling between the change of
length and stress enters only at $O(\Pero^2)$.

\subsection{Shear flow}

\begin{figure}
    \includegraphics[width=\textwidth]{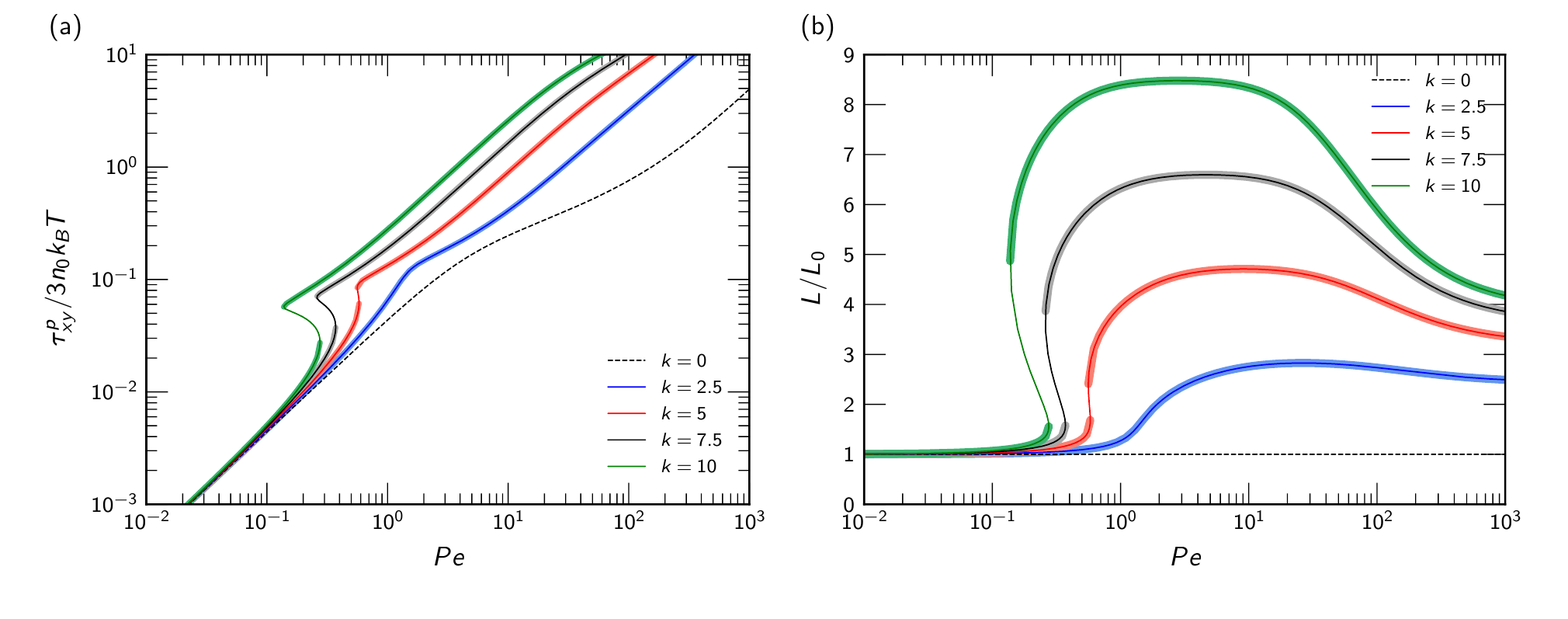}
    \caption{\label{fig:SSS-tauxy-L} 
    (a) Shear stress \emph{vs.}~dimensionless shear rate for different values of growth
    rate $k$. 
    (b) Normalized rod length in shear flow \emph{vs.}~dimensionless shear rate for
    different values of growth rate $k$, with model parameters $\alpha = 5$ and
$\beta = 500$.  The thick lines represent values accessible under shear rate
control; the entire curves are accessible under shear stress control. \SD{The black
dashed lines ($k=0$) show the values in the absence of any rod growth: i.e.~for a dilute suspension of rigid rods with constant length.}}
\end{figure}

\FiR{fig:SSS-tauxy-L}~(a) shows the shear stress as a function of $\Pero$ for
different values of $k$ and \FiR{fig:SSS-tauxy-L}~(b) shows the corresponding results
for rod length $L$.  Note that if $k = 0$ there is no growth of rods (as shown
by the black dashed line in \FiR{fig:SSS-tauxy-L}~(b)), and RRM
reduces to a suspension of monodisperse rigid rods. For low values of $k$ (see
graph for $k = 2.5$), the shear stress increases monotonically with shear rate.
The results are identical if shear stress is increased and the shear rate
obtained as an output.  From \FiR{fig:SSS-tauxy-L}~(b) we see the increase in rod length
at $\Pero \approx 1$: i.e.~the formation of FIS.  This leads to an increase in
shear viscosity, i.e.~shear-thickening behavior, as shown in
\FiR{fig:SSS-eta_k}. Note that at such low values of $k$, the viscosity is still
continuous. At sufficiently high $\Pero$, the rods break down, resulting in
shear-thinning.
\begin{figure}
    \includegraphics{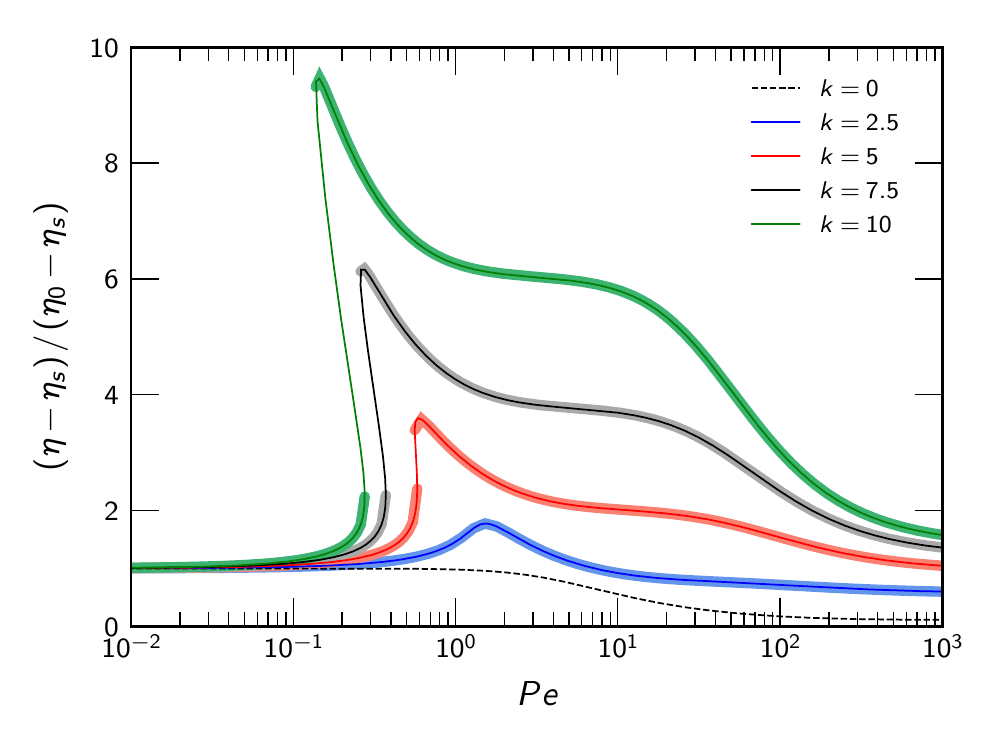}
    \caption{\label{fig:SSS-eta_k} Micelle contribution to shear viscosity normalized 
    by that at vanishing shear rate ($\eta_s$ is the solvent viscosity and 
    $\eta_0$ is the zero-shear viscosity of the solution) \emph{vs.}~dimensionless shear rate
    for different values of growth rate $k$, corresponding to the data presented
    in \FiR{fig:SSS-tauxy-L}. The thick lines represent values accessible under shear rate
    control; the entire curves are accessible under shear stress control.
    The black dashed line ($k=0$) shows the well-known shear-thinning result for
    rigid rod suspensions without reaction.}
\end{figure}

At higher values of the growth rate $k$, the curve of $\tau^p_{xy}$ \emph{vs.}~$\Pero$
displays a region where it is multivalued: three values of shear stress are
possible for the same shear rate.  Thus in a shear rate controlled experiment,
\FiR{fig:SSS-tauxy-L}(a) should show a jump discontinuity in stress, the height
of the jump increasing with increase in $k$. At constant $\Pero$ the
intermediate branch of the $\tau^p_{xy}$ \emph{vs.}~$\Pero$ is unstable. The
discontinuity is more prominent in viscosity data, as can be seen in
\FiR{fig:SSS-eta_k}. Only in a stress-controlled experiment would the
intermediate solution branch be observed. Note that the multivalued nature of
the flow curves presented here is in contrast to that observed in shear banding
of entangled WMS, where the shear rate becomes multivalued over a certain range of shear
stress.

\begin{figure}[tbh]
    \includegraphics[width=\textwidth]{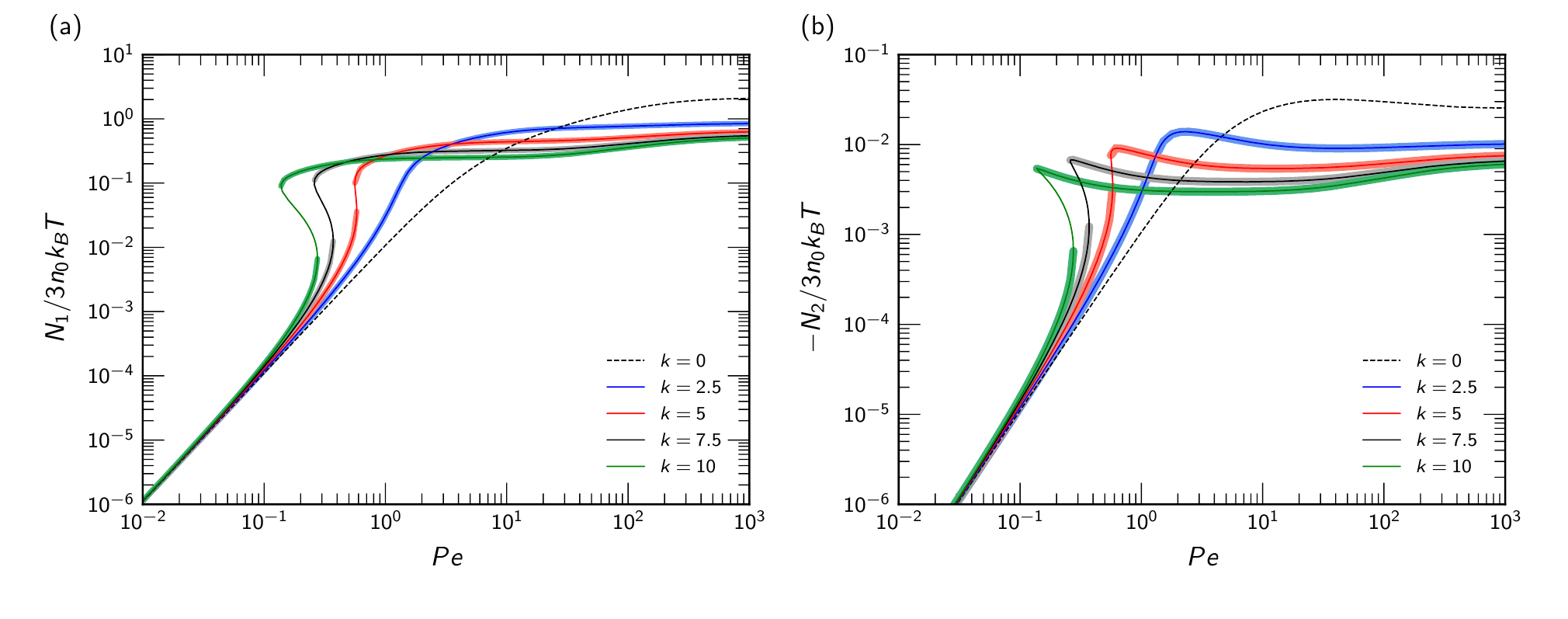}
    \caption{\label{fig:SSS-N1N2} First (a) and second (b) normal stress
    differences in shear flow \emph{vs.}~dimensionless shear rate for different values
    of growth rate $k$, with model parameters $\alpha = 5$ and $\beta = 500$.
    The thick lines represent values accessible under shear rate
control; the entire curves are accessible under shear stress control. \SD{The black
dashed lines ($k=0$) show the values in the absence of any rod growth.}}
\end{figure}

At higher values of $\Pero$, the multiplicity disappears -- we see from
\FiR{fig:SSS-eta_k} that this corresponds to  shear-thinning behavior.  The
multiplicity is also seen in the first and second normal stress differences
(\FiR{fig:SSS-N1N2}), though the second normal stress difference is numerically
much smaller.  The shear-thinning region is characterised by a decrease in rod
length (\FiR{fig:SSS-tauxy-L}~(b)) as $\Pero$ increases. 

\begin{figure}[tbh]
    \includegraphics{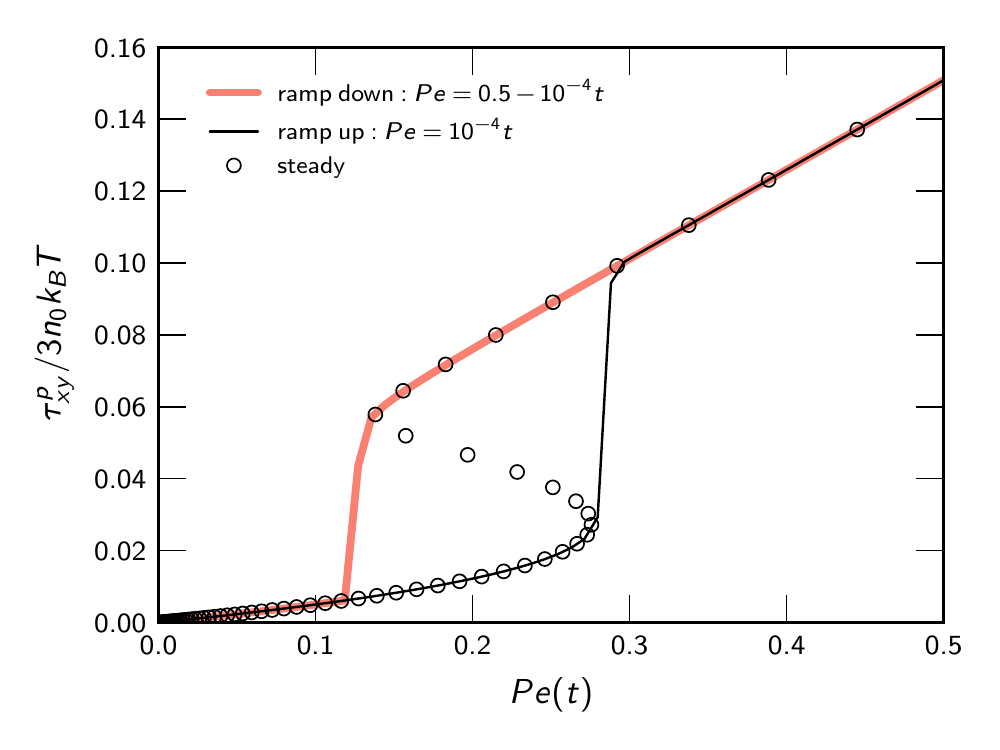}
    \caption{\label{fig:TSS_RMP} Transient shear stress \emph{vs.} $\Pero$ on ramping
    up (down) the shear rate through the region exhibiting multipicity in stress. 
    Model parameters are $k = 10$, $\alpha = 5$ and $\beta = 500$.
    Note the jumps between the branches on ramp up and ramp down indicating which
    specific value of stress is attained under transient conditions.}
\end{figure}
    
The presence of multiple values of stress at steady state for a given shear
rate obviously raises the question as to which specific value is attained
under transient conditions. To investigate this issue, we calculated the
transient shear stress for two cases -- (i) The shear rate is slowly ramped
up from zero through the multivalued region, and (ii) The shear rate is
slowly ramped down from beyond the multivalued region back to zero. \SD{To access the entire flow curve would require a quasisteady ramp in shear stress rather than shear rate.} For the
ramp calculations we define $\Pero(t) = 10^{-4} t$ on the upward ramp 
and $\Pero(t) =0.5-10^{-4}t$ on the downward, the ramps being deliberately
low so as to emulate a quasi-static process. The results are shown in
\FiR{fig:TSS_RMP}, along with the steady state data from
\FiR{fig:SSS-tauxy-L}(a) corresponding to $k = 10$. We find that on
ramping up, the transient stress follows the lower branch till the first turning
point, when it suddenly jumps up to the upper branch. The reverse is seen in
case of ramping down from beyond the multivalued region.

\begin{figure}[tbh]
    \includegraphics{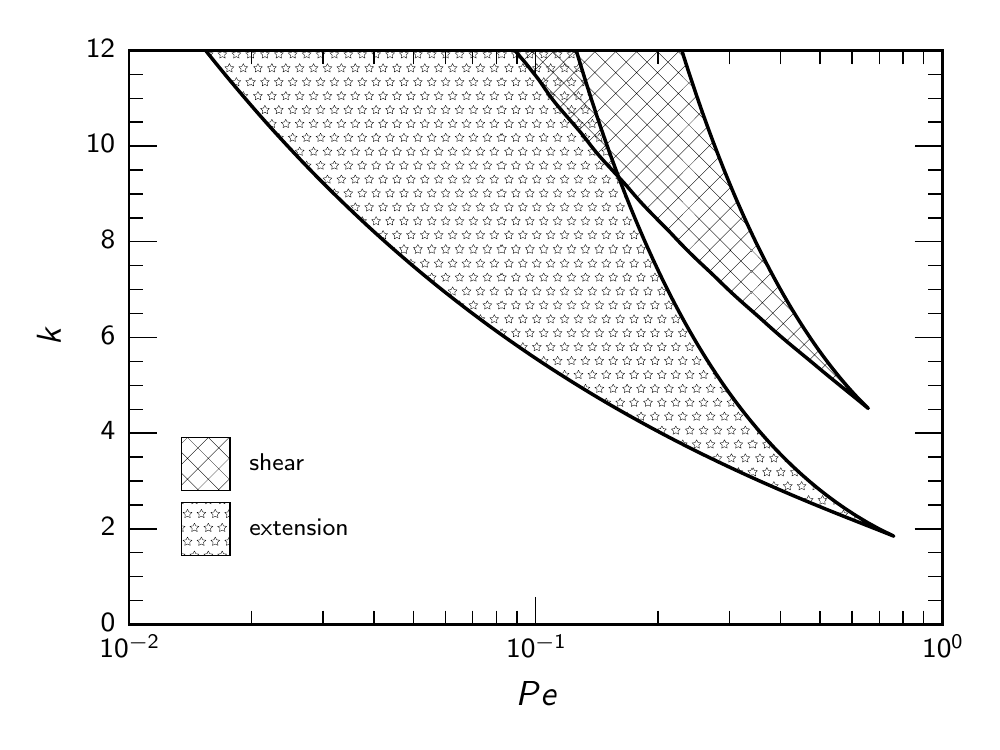}
    \caption{\label{fig:phase_k} Phase diagram showing the region in $k-\Pero$
    plane where the flow curve becomes multivalued.  Model parameters are $\alpha = 5$ and
    $\beta = 500$.}
\end{figure}

The region of multiplicity for both shear and extensional flow as a function of
$k$ and $\Pero$ is shown in \FiR{fig:phase_k}.  Note that this region is larger
for extension than for shear; in particular it extends to lower $k$. Recalling that $k$ is the rate constant for alignment-induced growth, this result makes sense physically because extension increases $\hat{S}$ more rapidly with $\Pero$ than does shear.

\begin{figure}[tbh]
    \includegraphics[width=\textwidth]{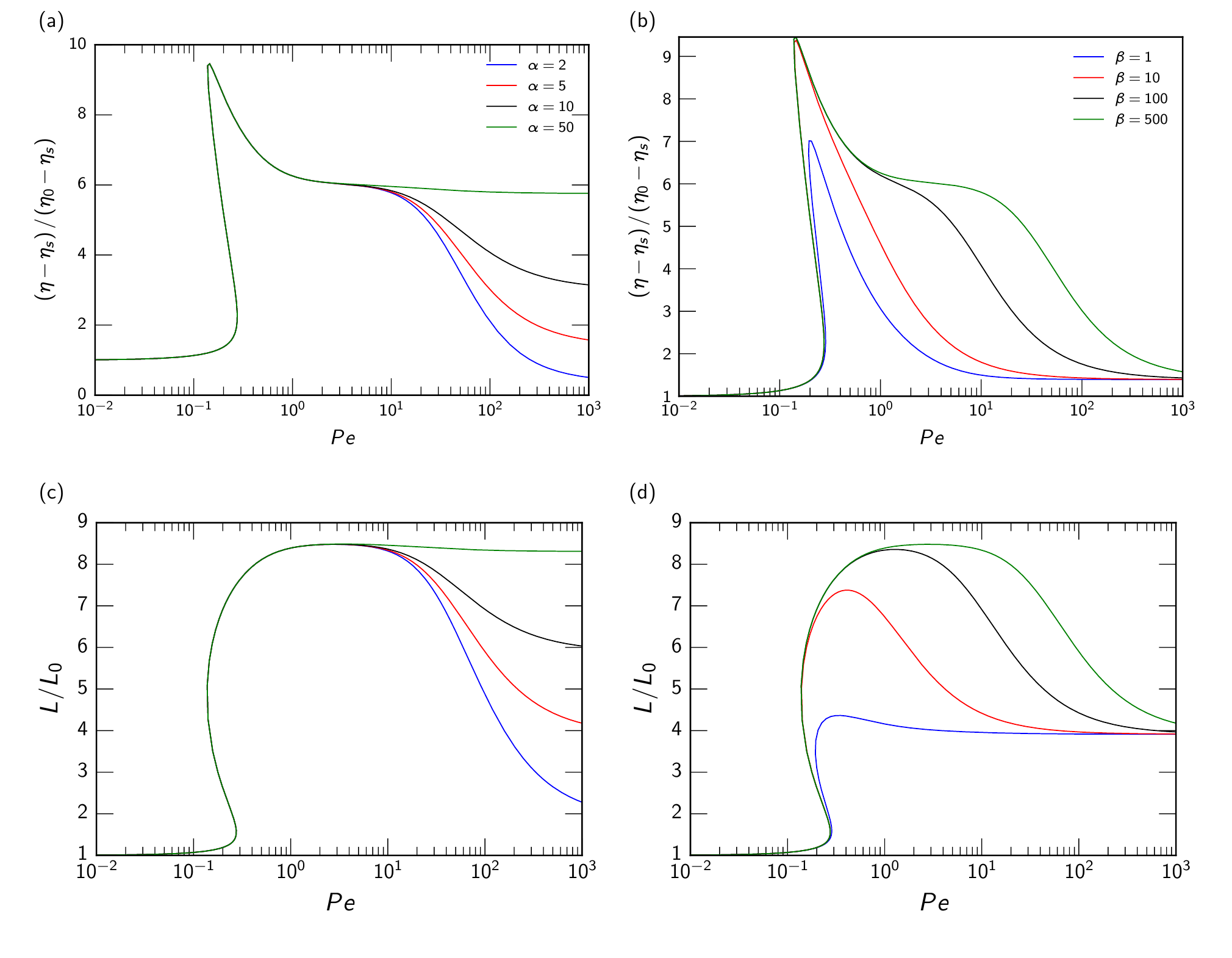}
    \caption{\label{fig:SSS-LaLb} Normalized shear viscosity \emph{vs.}~dimensionless shear
    rate for different values of (a) $\alpha$ with $\beta = 500$ and 
    (b) $\beta$ with $\alpha = 5$, and $k = 10$, under shear stress control.
    (c) and (d) show the dependence on rod length corresponding to (a) 
    and (b), respectively.}
\end{figure}

The effect of the parameters $\alpha$ and $\beta$ are shown in
\FiR{fig:SSS-LaLb}.  Recall that $\alpha$ and $\beta$ control the maximum rod
length at a given $\Pero$ (see \ER{eq:Lmax}). As expected, we see that $\alpha$
determines the rod length at very high $\Pero$, whereas $\beta$ determines the
shear rate where hydrodynamic stresses kick in to break down the rods. Note that
$\alpha$ and $\beta$ have minimal effect on the multivalued nature of the flow
curves, their primary role is to enforce shear-thinning behavior at high $\Pero$.

We compare the model predictions with experimental data for shear viscosity in
\FiR{fig:SSS-expt}. The datasets labeled $(L_1)$ and $(L_2)$ represent shear-rate
controlled data from Liu and Pine \cite{Liu1996} for an equimolar aqueous
solution of cetyltrimethylammonium bromide (CTAB) and sodium salicylate (NaSal)
at 500~ppm and 1000~ppm, respectively.  Dataset $(D)$ shows stress-controlled
data from Dehmoune et al \cite{Dehmoune2007} for octadecyltrimethylammonium
bromide (C$_{18}$TAB)/NaSal at 3~mM concentration.  For normalizing the viscosity
we have used the data reported at the lowest shear rate as the zero-shear
viscosity; the solvent viscosity (taking $\eta_s\approx 0.001$~Pas for water) is
negligible for all cases compared to the viscosity of the solution.  All the
three datasets were obtained from experiments performed in a Couette rheometer.
The model parameters in \FiR{fig:SSS-expt} were generated not through any
rigorous algorithm, but via simple trial-and-error and checked visually. Insofar as our present aim is to demonstrate that our model generates reasonable agreement
with experimental observations, we consider this to be sufficient.

While our model is able to capture the shear-thickening and the subsequent
shear-thinning behavior reasonably, the experimental data exhibits a small
amount of initial shear-thinning that is absent from the model predictions.
Note that a suspension of rigid rods is shear-thinning in the absence of
any reaction, but this is observed at a higher $\Pero$ compared to the onset
of the multivalued region (see \FiR{fig:SSS-eta_k} for $k = 0$).  The
problem here is primarily that polydispersity of rod lengths is neglected --
i.e.~the distribution is a $\delta$-function, and so is the associated time
scale. Such a distribution cannot be expected to capture features possibly
arising due to a broad distribution, as in an experimental system. An
analogous difficulty arises in using the FENE-P model for describing
polymer solutions.
We speculate that a certain amount
of flow-induced alignment is
required before the rods start reacting. Indeed, a small amount of initial
shear-thinning can be observed if the growth equation \ER{eq:Ra} is made to
depend on a higher power of $\hatS$, say the third power (a second power
produces negligible amount of shear-thinning). However, since our expression for
the growth rate is purely phenomenological and there is no specific reason for
choosing a particular power of $\hatS$, we have refrained altogether from
introducing yet another parameter for the power law.  Another issue worth
pointing out is that we have used the rotational diffusion coefficient $D_{r,0}$
as a fitting parameter. We find the fitted values to be significantly more than
their approximate physical values. As our model stands, the effect of changing
$D_{r,0}$ is to shift and stretch the viscosity curves along the
$\dot{\gamma}$-axis.  Nevertheless, we reiterate that our model predicts the
correct qualitative dependence of the shear-viscosity on the shear rate.

\begin{figure}[tbh]
    \includegraphics{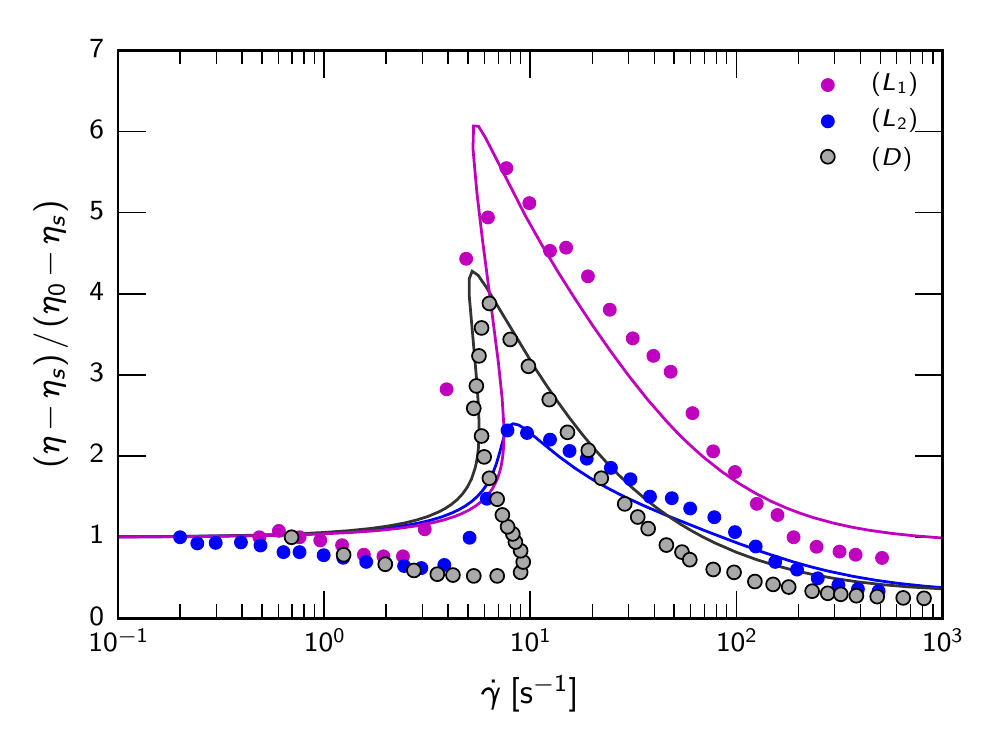}
    \caption{\label{fig:SSS-expt} Comparison of steady shear viscosity predicted
    by our model with experimental measurements.  $(L_1)$ and $(L_2)$ show data
    from Liu and Pine \cite{Liu1996}, and $(D)$ shows data from Dehmoune et at
    \cite{Dehmoune2007}.  Solid lines show results from our model fitted to the
    experimental measurements.  The values of the fitting parameters are 
    $(L_1)$: $D_{r,0} = 20\mathrm{s^{-1}}$, $k = 7.5$, $\alpha = 4$, $\beta = 10$;
    $(L_2)$: $D_{r,0} = 8\mathrm{s^{-1}}$, $k = 3.5$, $\alpha = 2$, $\beta = 50$;
    $(D)$: $D_{r,0} = 11.8\mathrm{s^{-1}}$, $k = 6$, $\alpha = 2$, $\beta = 6$.
}
\end{figure}

\subsection{Extensional flow}

\begin{figure}[tbh]
    \includegraphics[width=\textwidth]{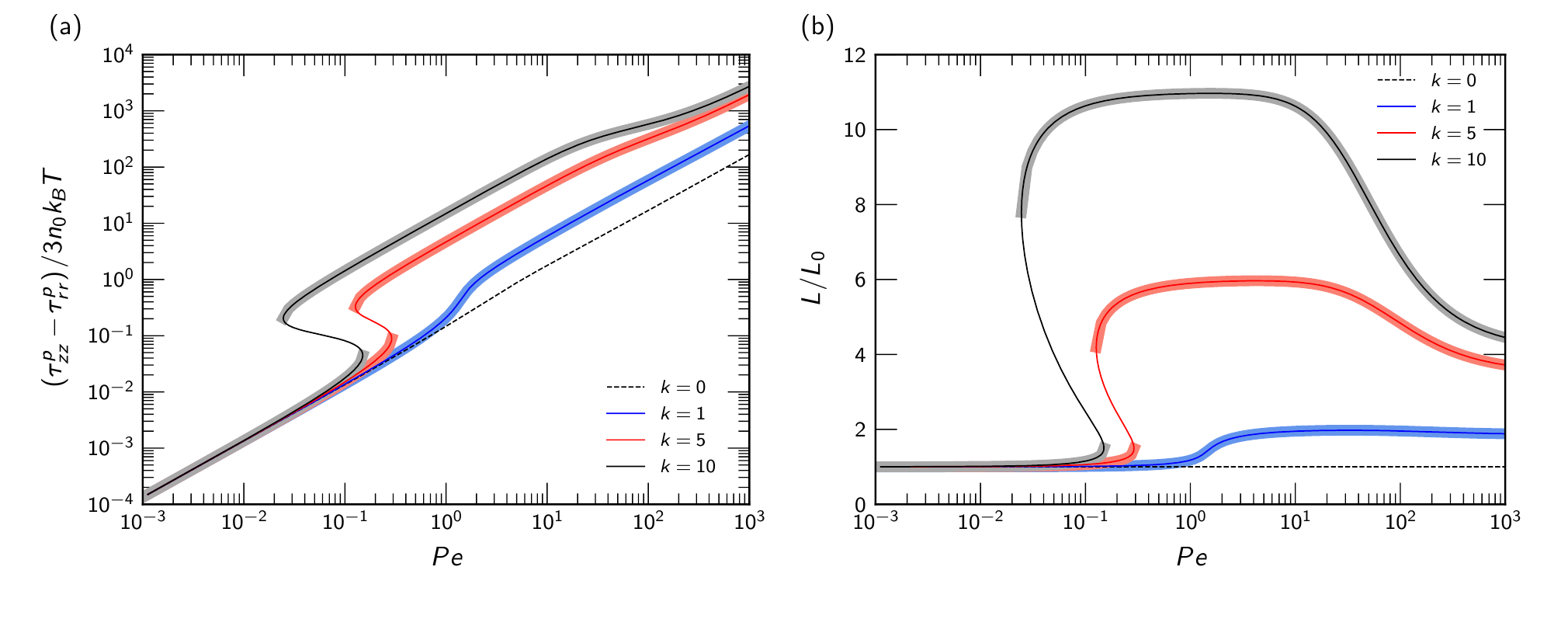}
    \caption{\label{fig:SUE-N1-L} (a) Extensional stress difference and
    (b) rod length \emph{vs.}~dimensionless strain rate at steady state in
    uniaxial extensional flow for different values of growth rate $k$, with
    model parameters $\alpha = 5$ and $\beta = 500$.  The thick lines represent
    values accessible under extension rate control; the entire curves are
accessible under stress control. \SD{The black
dashed lines ($k=0$) show the values in the absence of any rod growth.}}
\end{figure}

\begin{figure}[tbh]
    \includegraphics{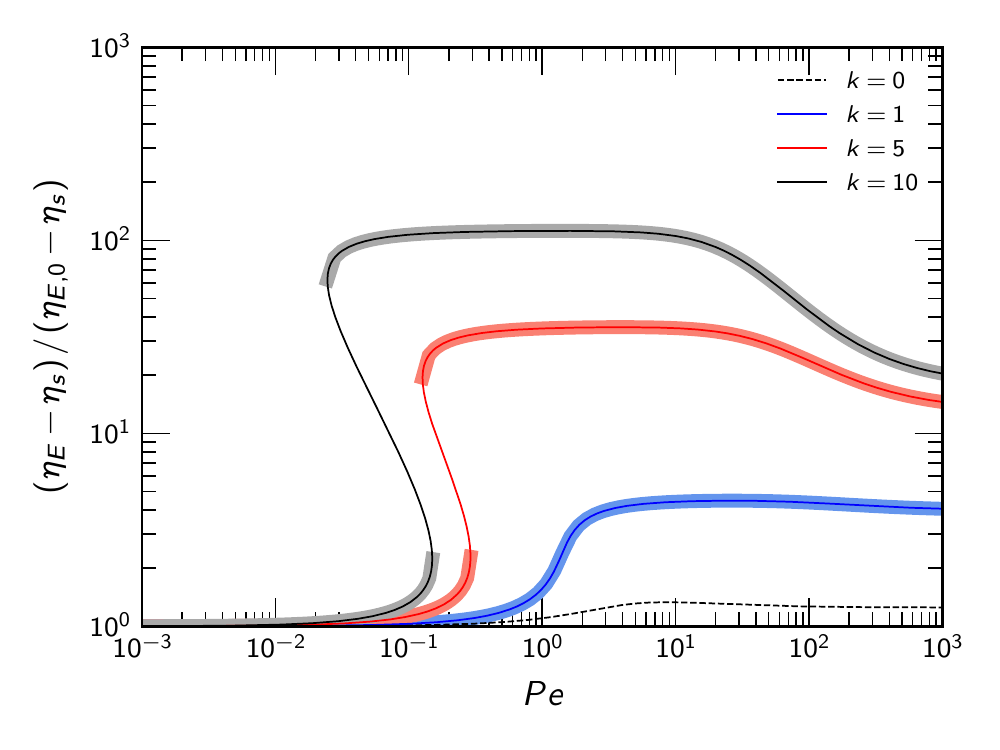}
    \caption{\label{fig:SUE-etaE_k} Extensional viscosity
    \emph{vs.}~dimensionless strain rate at steady state in uniaxial extensional
    flow for different values of growth rate $k$, corresponding to the data
    presented in \FiR{fig:SUE-N1-L}. \SD{The black
dashed lines ($k=0$) show the values in the absence of any rod growth.}}  
\end{figure}

\FiR{fig:SUE-N1-L}~(a) shows the extensional stress difference
$\tau^p_{zz}-\tau^p_{rr}$ as a function of $\Pero$ for different values of the
growth rate $k$ and \FiR{fig:SUE-N1-L}~(b) the corresponding plot for $L$.
Analogous to the situation in shear flow, there is flow-induced structure
formation and multiplicity in the stress response at higher values of $k$.  The
second normal stress difference is identically zero in uniaxial extension.  The
dependence of extensional viscosity $\eta_E = (\tau_{zz} - \tau_{rr})/\epdot$ is
shown in \FiR{fig:SUE-etaE_k}.  Compared to shear flow, the increase in rod
length resulting in extension-thickening is much higher. Correspondingly, the
region in parameter space exhibiting multiplicity is substantially larger for
extension than for shear as illustrated in \FiR{fig:phase_k}. 

\begin{figure}[tbh]
    \includegraphics{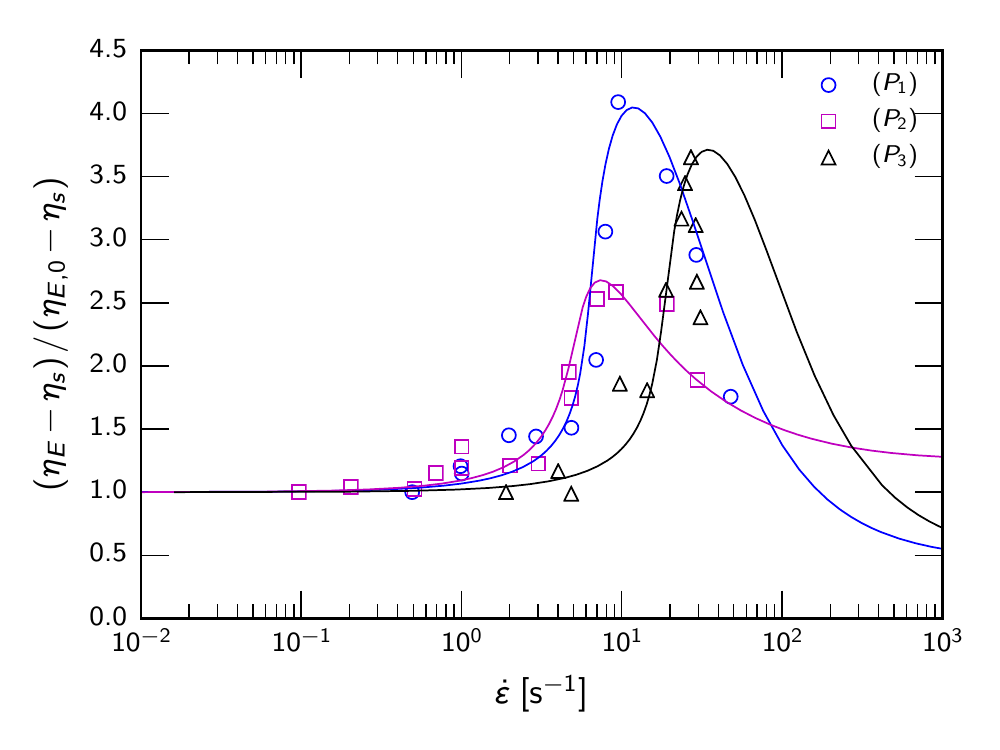}
    \caption{\label{fig:SUE-expt} Comparison of steady extensional
    viscosity with experimental data of Prud'homme and Warr
    \cite{Prudhomme1994}.  The solid lines show model predictions fitted to the
    experimental data.  The parameter values are 
    $(P_1)$: $D_{r,0} =  5.4\mathrm{s^{-1}}$, $k = 1.25$, $\alpha = 0.5$, $\beta = 10$;
    $(P_2)$: $D_{r,0} =  8.5\mathrm{s^{-1}}$, $k = 3.2$, $\alpha = 1$, $\beta = 1$;
    $(P_3)$: $D_{r,0} = 14.5\mathrm{s^{-1}}$, $k = 1.15$, $\alpha = 0.5$, $\beta = 10$.
}
\end{figure}

While we are not aware of experimental data reporting a multivalued extensional
viscosity, viscosity data showing dramatic extension-thickening has been
reported by Prud'homme and Warr \cite{Prudhomme1994} for equimolar solutions of
tetramethyltriammonium bromide (TTABr) and NaSal using an opposing jet
rheometer.  We compare our model predictions with their viscosity data in
\FiR{fig:SUE-expt}. The datasets labeled $(P_1)$ and $(P_2)$ show experimental
results for a 25.2~mM solution for nozzle diameters (of the opposing jet
rheometer) 2~mm/2~mm and 4~mm/4~mm, respectively; whereas dataset $(P_3)$ is for
a 69.2~mM solution for diameters 2~mm/2~mm.  Even though there is significant
scatter in the experimental data, the qualitative trend in terms of
shear-thickening matches reasonably well with the model predictions. However,
the issue with the large fitted values of $D_{r,0}$ alluded to for shear flow
holds here as well. Note that these experiments did not report a discontinuous
jump in viscosity, so the fits with the model are only for lower values of $k$. 

\section{Conclusion}

We presented a closed-form mechanistic model that can capture shear
(extension)-thickening and subsequent shear (extension)-thinning in dilute
wormlike micellar solutions. One of our goals in constructing the model was to
ensure that it is expressed in tensorial form, allowing it to be applied for
arbitrary flow fields, and that it is simple enough so that it can be
incorporated as a constitutive relation in existing solvers for numerical
simulation of complex flows, for example in investigating turbulent drag
reduction of micellar solutions.

Our model assumes micelles as a collection of rigid rods that can grow in
length in response to flow-induced alignment. There is a maximum length,
depending on shear (extension)-rate to which the rods may grow before they break
down into smaller rods. This mechanism is intended to capture rupture of long
micelles due to hydrodynamic stresses. We do not specify the microscopic details
of the growth or rupture processes, nor is such a specification necessary for
the intended  level of detail. The model couples the equations for time
evolution of the orientation tensor with a kinetic equation for rod length. The
equation for rod length has a growth term that is proportional to orientation
and a destruction term for capturing breakdown of rods.  For small shear
(extension) rates, our model predicts Newtonian behavior, whereas at relatively
higher values  shear (extension)-thickening is observed.  Furthermore, at
sufficiently high deformation rates, shear (extension)-thinning is observed. The
shear (extension)-thickening predicted corresponds to flow-induced structure
formation observed in experiments.  If the growth rate is increased beyond a
certain value, reentrant behavior (multiplicity of stress for a given
deformation rate) is observed.  Our model predictions agree reasonably well with
experimental data for both stress-controlled and shear-rate controlled cases. 


We believe that the work presented here takes an important step toward
understanding how to construct models for an important class of
self-assembling fluids, dilute wormlike micelle solutions. Such models will
have a broad impact on our capability to make reliable predictions about the
wide range of flow processes that involve such materials. Natural
applications for the present model include inhomogeneous flows, including
especially circular Couette flow, where many experiments have been
performed, as well as turbulent drag reduction, one of the most important
applications of these fluids. Finally, many improvements to the current
model can be envisioned. For example, more refined and fundamentally-based
models for the growth and breakage rates should be developed. Additionally,
the model assumes diluteness of the solution, which requires that $nL^3\ll
1$. In terms of our model $nL^3=n_0L_0 L^2$ which does not remain small when
$L$ is sufficiently large. Thus we overpredict the rotational diffusivity at
high $L$, where the solution effectively becomes semidilute. It would also
be desirable to overcome the assumption of a delta-function length
distribution without adding undue complexity.

%
\section*{Acknowledgments}
This material is based on work supported by the American Chemical Society
Petroleum Research Fund under grant number 53466-ND9 and the National Science
Foundation under grant number CBET-1604767.

%
\section*{References}
\nolinenumbers
\bibliography{refWMS}

\end{document}